\documentclass[12pt]{article}
\usepackage{graphicx}
\usepackage{cite}
\usepackage{color}
\newcommand{\abs}[1]{\left| #1\right|}

\newcommand{\x}[1]{{\textstyle #1}}
\newcommand{\xrm}[1]{{\textstyle \mbox{\rm #1}}}

\begin{document}
\title{Interference effects in the $X(4260)$ signal}
\author{
Eef~van~Beveren$^{\; 1}$ and George~Rupp$^{\; 2}$\\ [10pt]
$^{1}${\small\it Centro de F\'{\i}sica Computacional,
Departamento de F\'{\i}sica,}\\
{\small\it Universidade de Coimbra, P-3004-516 Coimbra, Portugal}\\
{\small\it eef@teor.fis.uc.pt}\\ [10pt]
$^{2}${\small\it Centro de F\'{\i}sica das Interac\c{c}\~{o}es Fundamentais,}\\
{\small\it Instituto Superior T\'{e}cnico, Universidade T\'{e}cnica de
Lisboa,}\\
{\small\it Edif\'{\i}cio Ci\^{e}ncia, P-1049-001 Lisboa, Portugal}\\
{\small\it george@ist.utl.pt}\\ [.3cm]
{\small PACS number(s): 14.40.Gx, 13.66.Bc, 14.40.Lb, 14.40.Cs}
}


\maketitle
\begin{abstract}
We show that the long known puzzling branching ratios of open-charm decays
in $e^+e^-$ annihilation can be reasonably described with a simple form factor,
which strongly suppresses open channels far above threshold. Application
to the $e^{+}e^{-}\to J/\psi\,\pi\pi$ data on the $X(4260)$ enhancement
recently reported by the BABAR Collaboration \cite{ARXIV08081543} allows a
good fit with a simple nonresonant cusp structure around the
$D_s^\ast D_s^\ast$ threshold. Moreover, we argue that a closer
look at the data reveals an oscillatory pattern, which we model as an
interference effect between a fast --- OZI-allowed --- and a slow ---
OZI-forbidden --- $J/\psi\,f_0(980)$ mode. Other candidates for similar
nonresonant enhancements are discussed.
\end{abstract}

Recently, the BABAR Collaboration presented new preliminary data for the
reaction $e^{+}e^{-}\to J/\psi\,\pi^{+}\pi^{-}$ \cite{ARXIV08081543}.
These data exhibit a much more pronounced peak in the $X(4260)$ \cite{PLB667p1}
region than in the first BABAR observation of this structure
\cite{PRL95p142001}, besides a rather constant signal for the remaining
invariant masses \cite{ARXIV09044351}. The new experimental analysis of the
$X(4260)$ using a nonrelativistic Breit-Wigner parametrization yielded a mass
of $M=\left( 4252\pm 6^{+2}_{-3}\right)$ MeV and a width of
$\Gamma =\left( 105\pm 18^{+4}_{-6}\right)$ MeV.

The $X(4260)$ enhancement was confirmed, and also seen in the processes
$\pi^{0}\pi^{0}J/\psi$ as well as $K^{+}K^{-}J/\psi$, by the CLEO collaboration
\cite{PRL96p162003}, whereas the Belle Collaboration observed a similar
structure in $J/\psi\,\pi^{+}\pi^{-}$ \cite{PRL99p182004}.
On the theoretical side \cite{NPPS170p248,NPPS187p145,PLB625p212},
a variety of model explanations have been suggested, such as
a standard vector charmonium state ($4S$)
\cite{PRD72p031503},
a mesonic or baryonic molecule
\cite{PRD72p054023},
a gluonic excitation (hybrid)
\cite{PLB631p164},
or a $cq\bar{c}\bar{q}$ state
\cite{PRD72p031502}.

It was also noticed \cite{HEPEX0701002}
that the Belle data \cite{PRL99p182004}
reveal a curious $\pi\pi$ mass spectrum
in the $m_{\pi^+\pi^-J/\psi}$ invariant-mass region of 4.2--4.4 GeV,
which was confirmed by the BABAR Collaboration
in Ref.~\cite{ARXIV08081543}.

A remarkable aspect of this experimental observation
is that the main signal of $e^{+}e^{-}\to\pi^{+}\pi^{-}J/\psi$
coincides with the $D_{s}^{\ast}D_{s}^{\ast}$ threshold
\cite{ARXIV09044351}, and furthermore that
very little $D\bar{D}$ production has been observed in the 4.26 GeV region
\cite{PRD76p111105}.  Here, we shall first pay some attention to the latter
phenomenon, then discuss the relation of the $X(4260)$ signal
to the opening of $D_{s}^{\ast}D_{s}^{\ast}$, finally
and present a possible explanation for the $\pi\pi$ mass spectrum.

It has been observed \cite{PRD76p111105,ARXIV07100165}
that $D\bar{D}$ production in $e^{+}e^{-}$ annihilation
is suppressed at invariant masses far above
the production threshold for $D\bar{D}$.
Branching fractions have been measured at 4.028 GeV
with the SLAC/LBL magnetic detector at SPEAR \cite{PLB69p503}.
The results suggest that the opening of a channel
is followed by a rather fast fading out of the same channel
at higher invariant masses.
In Refs.~\cite{NPB471p59,ZPC68p647}, it was shown that this feature can be
parametrized for the total cross section $\sigma$
by a simpel Gaussian form factor, which for $P$-waves suggests
the expression
\begin{equation}
\sigma\propto \abs{pr_{0}}\, e^\x{-\abs{pr_{0}}^{2}}
\;\;\; ,
\label{crossdamp}
\end{equation}
where $p$ stands for the two-particle linear momentum
and $r_{0}$ represents a distance parameter.
In Table~\ref{D0D0branching}, we show some results
of formula~(\ref{crossdamp}) for the branching fractions
of pairs of neutral charmed mesons at 4.028 GeV, produced in
$e^+e^-$ annihilation.
\begin{table}[ht]
\begin{center}
\begin{tabular}{||c||c|c|c||}
\hline\hline & & & \\ [-10pt]
channels & $r_{0}=3.2$ GeV$^{-1}$ & $r_{0}=3.8$ GeV$^{-1}$ &
experiment \cite{PLB69p503} \\
\hline\hline & & & \\ [-10pt]
$D^{0}D^{0}/D^{0}D^{\ast 0}$ & 0.021 & 0.007 & 0.05$\pm$0.035\\ [10pt]
$D^{0}D^{0}/D^{\ast 0}D^{\ast 0}$ & 0.0023 & 0.0002 &
0.0016$\pm$0.0014\\ [10pt]
$D^{0}D^{\ast 0}/D^{\ast 0}D^{\ast 0}$ & 0.11 & 0.033 & 0.031$\pm$0.016\\
\hline\hline
\end{tabular}
\end{center}
\caption[]{\small
Branching fractions for $e^{+}e^{-}\to D^{0}D^{0}$,
$D^{0}D^{\ast 0}$, and $D^{\ast 0}D^{\ast 0}$ at 4.028 GeV,
obtained by the use of formula (\ref{crossdamp}),
including the ratios which stem from the three-meson verticex
couplings, viz.\ $PP:PV:VV=1:4:7$
($P=$pseudoscalar, $V=$vector meson),
and compared to the results of Goldhaber {\it et al.} \cite{PLB69p503}.
}
\label{D0D0branching}
\end{table}
We observe that expression~(\ref{crossdamp})
allows a reasonable agreement with experiment.
This may explain why so few $D\bar{D}$ pairs are observed
at 4.26 GeV \cite{PRD76p111105}, as this channel opens at 3.74 GeV.
It supports the idea that channels get effectively
damped at invariant masses far above their thresholds. This phenomenon
may be understood as the manifestation of boost effects on the
wave functions of the produced mesons \cite{ZPC13p43}.
Consequently, near the $X(4260)$ we may restrict ourselves
to the opening of the $D_{s}^{\ast}D_{s}^{\ast}$ channel.
As for the possible alternatives in this energy region, the
$DD_{1}(2420)$ \cite{PRD79p014001} threshold at about 4.29 GeV lies
somewhat too high, while the $DD_{1}(2430)$ channel, involving the very
broad $D_1(2430)$ \cite{PLB667p1} resonance, cannot give rise to
the rather sharp $X(4260)$ enhancement either. Also the suggested
\cite{PRD79p014001} $D^*D_0^*(2400)$ threshold at roughly 4.3--4.4 GeV
\cite{PLB667p1}, with the very broad scalar charm meson $D_0^*$, is way
too smeared out to produce a narrow signal at about 4.25 GeV

In Refs.~\cite{ARXIV08111755,ARXIV09044351},
it was observed that $c\bar{c}$ resonances show up as dips
in the $e^{+}e^{-}\to\pi^{+}\pi^{-}J/\psi$ cross section,
and not as resonance peaks.
Furthermore, in Ref.~\cite{ARXIV09044351}
we showed that also at the opening of channels,
in particular open-charm baryonic channels,
dips appear in the $e^{+}e^{-}\to\pi^{+}\pi^{-}J/\psi$ cross section.
\clearpage

\begin{figure}[htbp]
\begin{center}
\begin{tabular}{c}
\scalebox{1.0}{\includegraphics{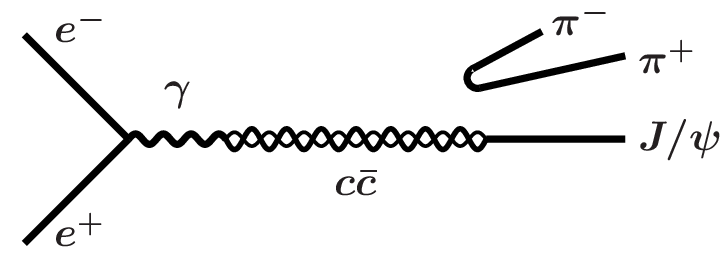}}\\ [-15pt]
\end{tabular}
\end{center}
\caption[]{\small
Peripheral creation of pion pairs in the gluon cloud surrounding
$c\bar{c}$, for the reaction $e^{+}e^{-}\to\pi^{+}\pi^{-}J/\psi$.
}
\label{Jpsipipi}
\end{figure}

In fact, the overall aspect of the $J/\psi\,\pi\pi$ cross section appears
to be rather constant (see below), with no obvious sign of the
established and possible new $c\bar{c}$ resonances.
So pion-pair creation does not seem to stem from the constituent $c\bar{c}$
system.  Nevertheless, besides a $c\bar{c}$ component we may assume
the presence of glue \cite{PRD56p4062,BJP34p865}.
The latter field will, in the periphery, absorp little of the $c\bar{c}$
oscillations and the corresponding resonances.
Hence, anything created out of the surrounding glue will probably not display
much of the charmonium structure and its spectrum. Thus, we may suppose that
the pion pair stems from the glue, not from the strongly oscillating interior.
This explains why a $\sigma$-like $\pi\pi$ structure can be formed that
is not correlated with any $c\bar{c}$ resonances. Such a structure, being very
broad, allows for a wide range of total two-pion masses at a slowly varying
rate, and so shows up as an almost constant signal for a comparably wide range
of total $J/\psi\pi\pi$ invariant masses. The peripheral, OZI-forbidden
$\pi\pi$-creation process is depicted in Fig.~\ref{Jpsipipi}.

However, at the opening of an open-charm decay channel,
the dynamics of the system is dominated by string breaking
through the creation of light quark-antiquark pairs.
We have shown in Ref.~\cite{ARXIV09044351}
that such a process is substantially faster
than peripheral pion-pair production.
Hence, it eats away signal by premature decay into open-charm hadrons,
thus leaving dips in the production cross section for $J/\psi\,\pi\pi$.
Nonetheless, this is not the case at the $X(4260)$ enhancement, which
certainly calls for an explanation.

In Ref.~\cite{ARXIV09044351}, we found
that, except for the dips and the $X(4260)$ signal,
the $e^{+}e^{-}\to\pi^{+}\pi^{-}J/\psi$ cross section
is rather flat.
This leaves us with the picture that the $c\bar{c}$ propagator,
which is formed in $e^{+}e^{-}$ annihilation,
allows for the development of a pion pair as long as it
is not resonating or near the opening of a threshold.
Moreover, as the $c\bar{c}$ propagator dominantly
couples to vector-vector (VV) open-charm pairs,
four times more weakly to pseudoscalar-vector (PV),
and seven times more weakly to pseudoscalar-pseudoscalar (PP),
it has to be expected that special phenomena may be observed at
or just above the threshold of a VV open-charm pair.
As we have argued above, open channels get damped quite fast at higher
energies.

There are two candidates for VV open-charm channels, viz.\
$D^{\ast}D^{\ast}$ and $D_{s}^{\ast}D_{s}^{\ast}$.
Now, $D^{\ast}D^{\ast}$ decay results from the creation
of $u\bar{u}$ and $d\bar{d}$ pairs, which, being a much faster process
than peripheral pion-pair creation, eats away signal from
$e^{+}e^{-}\to\pi^{+}\pi^{-}J/\psi$. So this might leave
a dip in the cross section of this process
\cite{ARXIV09044351}.
We shall come back to this issue further on.

We now assume for the reaction $e^{+}e^{-}\to\pi^{+}\pi^{-}J/\psi$
the following scenario.
When not near a resonance or an open-charm threshold,
the pion pair is basically formed \cite{JPG14p1037}
in the gluon cloud, with the quantum numbers of the $\sigma$.
However, near the $c\bar{c}\to D_{s}^{\ast}D_{s}^{\ast}$
threshold, the dynamics of the system is dominated
by $s\bar{s}$ pair creation, which we reckon to be
a much faster process than peripheral pion-pair creation.
Consequently, we should expect a dip in the cross section.
In order to explain the $X(4260)$ peak, we must assume that
$s\bar{s}$ pair creation, besides allowing for
the formation of the pair of $D^{\ast}_{s}$ mesons,
makes a process possible other
than the creation of a pion pair from up and down quarks.
Such a process exists, namely the formation of $f_{0}(980)$
\cite{PLB495p300}, which then couples to a pion pair, though
not very strongly  \cite{NPB320p1}.

This may explain why the $J/\psi\pi\pi$ signal closely follows
the rise and fall of the $D^{\ast}_{s}\bar{D}^{\ast}_{s}$ amplitude
on top of the constant ``background'' of $J/\psi\pi\pi$,
which stems from the processes originating in the surrounding glue.

\begin{figure}[htbp]
\begin{center}
\begin{tabular}{c}
\scalebox{1.0}{\includegraphics{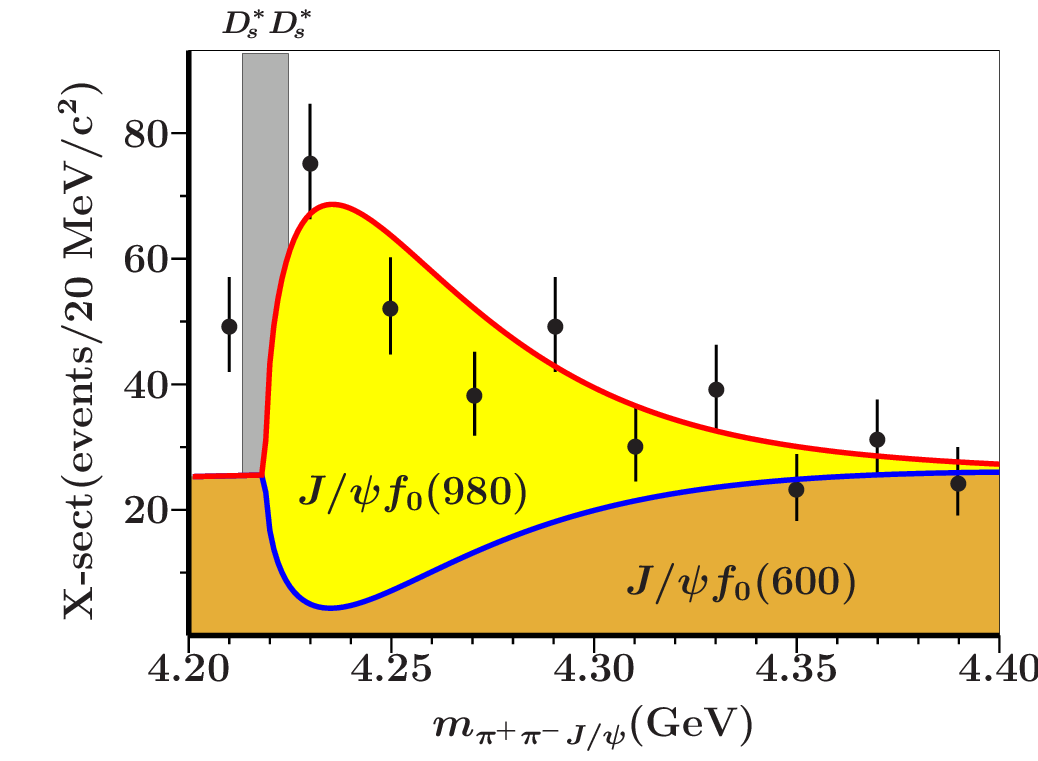}}\\ [-15pt]
\end{tabular}
\end{center}
\caption[]{\small
Schematic representation of
the distinct contributions to the total cross section
for $e^{+}e^{-}$ annihilation into $J/\psi\,\pi^{+}\pi^{-}$.
The vertical gray bar shows the region where
the $D_{s}^{\ast}D_{s}^{\ast}$ threshold may be situated,
according to the various results for the $D_{s}^{\ast}$ mass,
reported in
Refs.~\cite{PRL53p2465,PLB156p441,PRL58p2171,PLB207p349,PRD50p1884,PRL75p3232}.
The data are taken from Ref.~\cite{ARXIV08081543}.
}
\label{Jpsif0}
\end{figure}
In Fig.~\ref{Jpsif0} we depict each of the three contributions:
the almost constant peripheral production of pion pairs via sigmas,
the dip which is caused by premature decay into $D^{\ast}_{s}$ pairs,
and the contribution to pion-pair production by $f_{0}(980)$s
which stem from the abundantly produced $s\bar{s}$ pairs.
>From Fig.~\ref{Jpsif0} we may also infer that each
of the two distinct processes occurs for roughly 50\%.
Experiment \cite{PRL95p142001,PRL99p182004,ARXIV08081543}
seems to agree with this value,
but for a definite conclusion we must await better statistics.

We thus have two processes of very different origin, which result
in the same final state. Such a situation may give rise to interference
effects in the $J/\psi\,\pi\pi$ production amplitude, and hence to
observable oscillations, which indeed appear to be present in the data
shown in Fig.~\ref{Jpsif0}.
One mode is the relatively slow, OZI-forbidden process
via the peripheral formation of up and down quarks, while
the other mode is $J/\psi\,\pi\pi$ production
through the formation of $f_{0}(980)$s in an $s\bar{s}$-rich environment.
Moreover, for the latter process we have an idea of
the frequency of the $c\bar{c}$ oscillations, namely $\omega\sim 200$ MeV
\cite{PRD27p1527,HEPPH0201006}, which is equivalent to an oscillation
period of $T\approx5$ GeV$^{-1}$.

In the foregoing, we have taken the mass
of the $D_{s}^{\ast}D_{s}^{\ast}$ threshold
as a very rigorous boundary for the onset of the $X(4260)$ signal.
This is, of course, correct for the onset of the dip.
However, for $f_{0}(980)$ formation we are not bound
by the $D_{s}^{\ast}D_{s}^{\ast}$ threshold,
since $s\bar{s}$ creation will certainly be important already
close to but below threshold, also because of the 40--100 MeV width
\cite{PLB667p1} of the $f_0(980)$.
The BABAR data indeed start to rise already
some 40--50 MeV below threshold.
Furthermore, the $J/\psi f_{0}$ system couples in an $S$-wave
to the $c\bar{c}$ vector propagator.
Hence, the behavior will be different from the $P$-wave shape
of Eq.~(\ref{crossdamp}).

At present, it is not possible to model such a highly complex system,
since moreover the $c\bar{c}$ resonances
will play an important role \cite{ARXIV08091149}, too.
Hence, in order to account for interference,
we simply modify the $S$-wave equivalent
of the distribution in Eq.~\ref{crossdamp}
with an interference term:
\begin{equation}
\xrm{main signal}+
\left[1+\alpha\,
\cos\left(\left\{ m_{\pi^{+}\pi^{-}J/\psi}-2m_{D_{s}^{\ast}}\right\}
\Delta T\right)\right]
\; e^\x{-\abs{pr_{0}}^{2}}
\;\;\; ,
\label{decayamplitude}
\end{equation}
with $4p^{2}=m^{2}_{\pi^{+}\pi^{-}J/\psi}-4m^{2}_{D_{s}^{\ast}}$.
The main signal has been explained in Ref.~\cite{ARXIV09044351}.
It consists of a constant term and a very wide bell-shaped contribution,
which for the new BABAR data has its maximum at 4.35 GeV,
and a width of 750 MeV.
Here, we choose the constant contribution somewhat smaller
than in Ref.~\cite{ARXIV09044351},
in an attempt to average over the maxima and minima in the signal.

\begin{figure}[htbp]
\begin{center}
\begin{tabular}{cc}
\scalebox{0.78}{\includegraphics{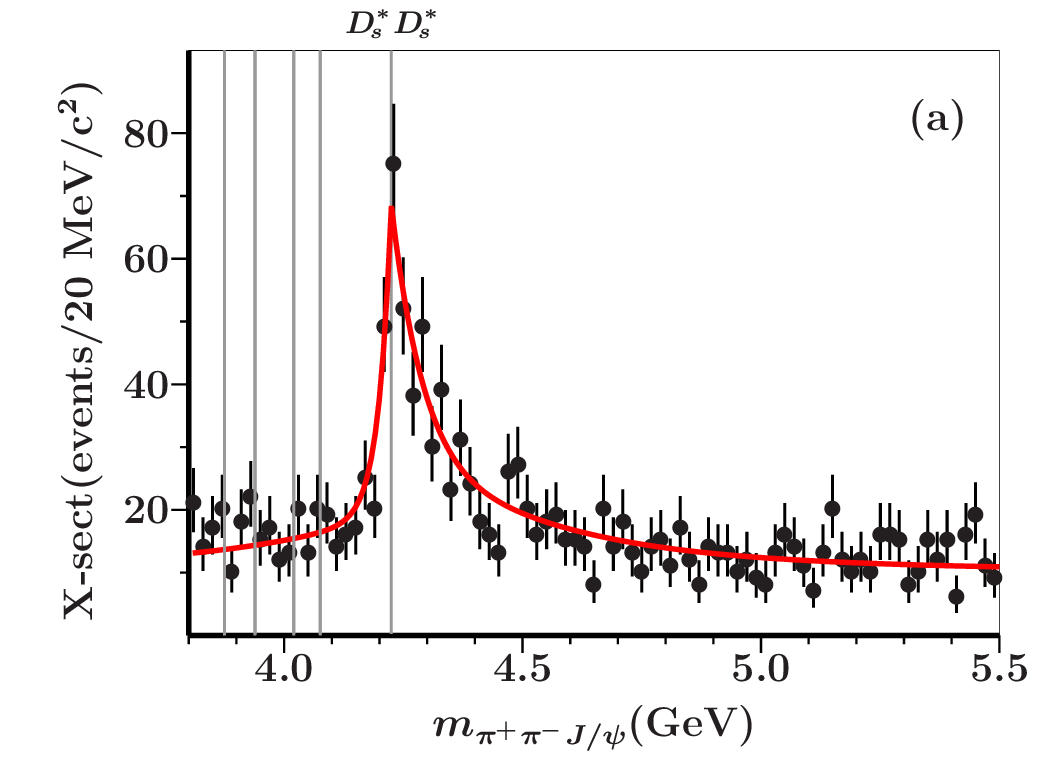}} &
\scalebox{0.78}{\includegraphics{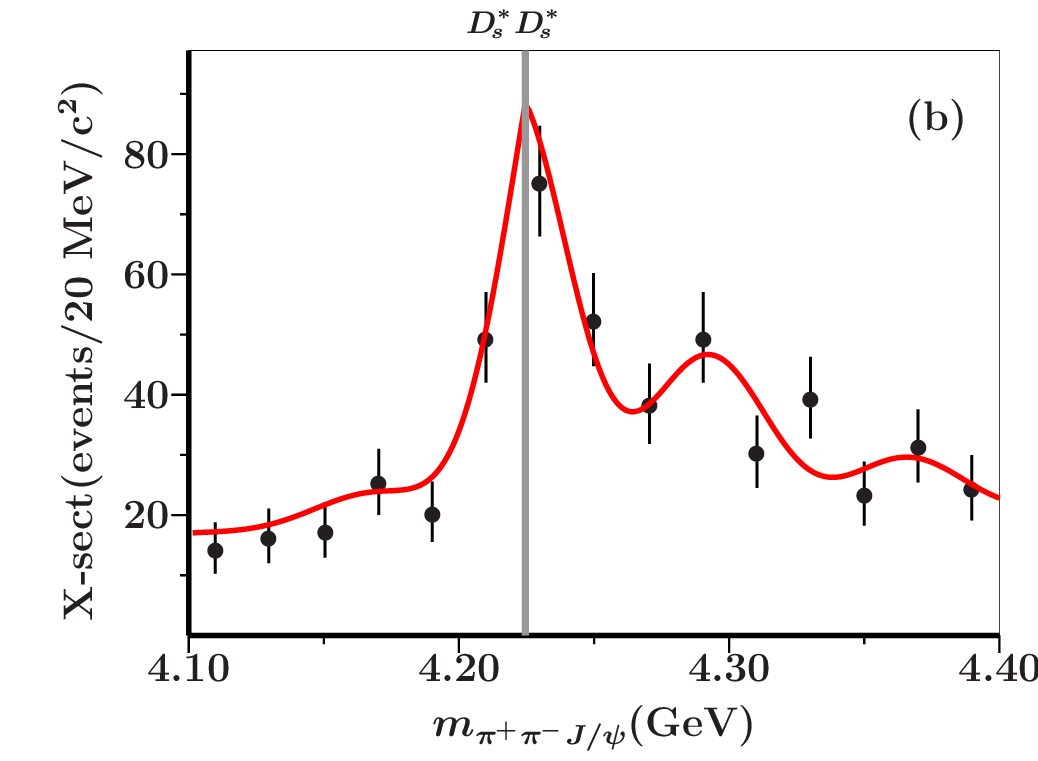}}\\ [-15pt]
\end{tabular}
\end{center}
\caption[]{\small
The new preliminary BABAR data~\cite{ARXIV08081543}
for $e^{+}e^{-}$ annihilation into $J/\psi\,\pi^{+}\pi^{-}$
in the invariant-mass region from 3.8 GeV to 5.5 GeV (a);
from 4.1 GeV to 4.4 GeV (b).
In (a), where no interference is considered,
we observe the $D_{s}^{\ast}D_{s}^{\ast}$ threshold
cusp in $J/\psi\,\pi\pi$.
In (b) an additional interference effect
is introduced as given by Eq.~(\ref{decayamplitude}), with
$\alpha =0.4$ and $\Delta T=85$ GeV$^{-1}$.
The vertical gray lines represent the various open-charm thresholds.
}
\label{interference}
\end{figure}

In Fig.~\ref{interference}a, we show the resulting amplitude over
a wide energy range, using $r_{0}=2.5$ GeV$^{-1}$ above threshold,
$r_{0}=4.2$ GeV$^{-1}$ below, and for the moment $\alpha =0$.
On the other hand, in Fig.~\ref{interference}b
we depict the amplitude in the energy interval of 4.1--4.4 GeV, but
now for  $\alpha =0.4$ and $\Delta T=85$ GeV$^{-1}$, and the same
values for $r_0$.
With this parametrization we clearly obtain
the expected $D_{s}^{\ast}D_{s}^{\ast}$ threshold cusp
\cite{JPG35p075005}
in the cross section for
$e^{+}e^{-}$ annihilation into $J/\psi\,\pi^{+}\pi^{-}$.
Also the interference pattern peaks at threshold
and agrees to a reasonable degree with the data.
Consequently, the BABAR data seem
to confirm our picture for the $X(4260)$ signal
as consisting of two processes, each with
its own characteristic frequency.
Furthermore, we find from expression~(\ref{decayamplitude})
that peripheral pion-pair creation oscillates
with a period of about 85 GeV$^{-1}$,
which is 17 times slower
than pion-pair production via $s\bar{s}$.
This looks like a reasonable factor for the suppression of OZI-forbidden
hadronic decays relative to OZI-allowed ones.
Also note that, except
at the opening of the $c\bar{c}\to D_{s}^{\ast}D_{s}^{\ast}$ channel,
there is no specific peak position associated
with either of the two phenomena.
Hence, in this scenario the $X(4260)$ enhancement
does not represent a new kind of resonance.

Let us now further discuss the $D^{\ast}D^{\ast}$ channel.
In this case, the relevant pair creations from string breaking are
$u\bar{u}$ and $d\bar{d}$, which then couple to the $\sigma$.
Consequently, the situation is comparable to the $X(4260)$ signal.
In fact, the Belle Collaboration did observe a similar structure
\cite{PRL99p182004}, just above the $D^{\ast}D^{\ast}$ threshold,
but not confirmed by the BABAR Collaboration \cite{ARXIV08081543}.
However, there is an important difference between the mass
of the latter signal and of the $X(4260)$.
Namely, the $X(4260)$ comes right in between the two $c\bar{c}$
vector states $\psi (4160)$ and $\psi (4415)$.
So the amplitude for open-charm production is close to a minimum
here, which is favorable for the observation of other phenomena,
as explained above. On the other hand,
the Belle signal at 4050 MeV comes almost on top
of the $\psi (4040)$, and so may be subject to destructive interference.
Nonetheless, this might be settled with better statistics in the future.

In Ref.~\cite{ARXIV09044351}, we showed that the cross section
for $e^{+}e^{-}$ annihilation into $J/\psi\,\pi^{+}\pi^{-}$
consists of a constant signal plus and a very wide bell-shaped structure,
which has its maximum at about 4.35 GeV. The latter extremely broad
structure is supported by a very recent detailed three-body calculation
for $J/\psi\,\pi^{+}\pi^{-}$ \cite{EKA}, to be published soon.
In the foregoing, we have shown that in the 4.26 GeV region there
additionally appears to be an interference structure.
We believe a 10 MeV binning of the data could shed some more light on
the picture for the $X(4260)$ enhancement proposed here.

The fundamental implications of the here observed interference effects
between equal final states but with different creation mechanisms are not
yet clear to us. Nevertheless, similar effects might be observable for
the $X(4140)$ enhancement in $J/\psi\,\phi$ \cite{ARXIV09032229}.

\vskip 7pt
We are grateful for the rather precise measurements
of the BABAR Collaboration, which allowed for the present analysis.
This work was supported in part by
the \emph{Funda\c{c}\~{a}o para a Ci\^{e}ncia e a Tecnologia}
\/of the \emph{Minist\'{e}rio da Ci\^{e}ncia, Tecnologia e Ensino Superior}
\/of Portugal, under
contract POCI/\-FP/\-81913/\-2007.

\newcommand{\pubprt}[4]{#1 {\bf #2}, #3 (#4)}
\newcommand{\ertbid}[4]{[Erratum-ibid.~#1 {\bf #2}, #3 (#4)]}
\def\BJP{Braz.\ J.\ Phys.}
\def\JPG{J.\ Phys.\ G}
\def\NPB{Nucl.\ Phys.\ B}
\def\NPPS{Nucl.\ Phys.\ Proc.\ Suppl.}
\def\PLB{Phys.\ Lett.\ B}
\def\PRD{Phys.\ Rev.\ D}
\def\PRL{Phys.\ Rev.\ Lett.}
\def\ZPC{Z.\ Phys.\ C}


\begin{thebibliography}{39}
\bibitem{ARXIV08081543}
B.~Aubert  [BaBar Collaboration],
{\it Study of the $\pi^{+}\pi^{-}J/\psi$ mass spectrum
via Initial-State Radiation at BaBar},
arXiv:0808.1543 [hep-ex].

\bibitem{PLB667p1}
C.~Amsler {\it et al.} \/[Particle Data Group Collaboration],
{\it Review of Particle Physics},
\pubprt{\PLB}{667}{1}{2008}.

\bibitem{PRL95p142001}
B.~Aubert {\it et al.}  [BABAR Collaboration],
{\it Observation of a broad structure in the $\pi^{+}\pi^{-}J/\psi$
mass spectrum around 4.26-GeV/c$^{2}$},
\pubprt{\PRL}{95}{142001}{2005}
[arXiv:hep-ex/0506081].

\bibitem{ARXIV09044351}
E.~van Beveren and G.~Rupp,
{\it The $X(4260)$ and possible confirmation
of $\psi(3D)$, $\psi(5S)$, $\psi(4D)$, $\psi(6S)$ and $\psi(5D)$
in $J/\psi\pi\pi$},
arXiv:0904.4351 [hep-ph].

\bibitem{PRL96p162003}
T.~E.~Coan {\it et al.}  [CLEO Collaboration],
{\it Charmonium decays of Y(4260), psi(4160), and psi(4040)},
\pubprt{\PRL}{96}{162003}{2006}
[arXiv:hep-ex/0602034].

\bibitem{PRL99p182004}
C.~Z.~Yuan {\it et al.}  [Belle Collaboration],
{\it Measurement of $e^{+}e^{-}\to\pi^{+}\pi^{-}J/\psi$ cross section
via initial-state radiation at Belle},
\pubprt{\PRL}{99}{182004}{2007}
[arXiv:0707.2541 [hep-ex]].

\bibitem{NPPS170p248}
B.~D.~Yabsley,
{\it New charmonium-like states},
\pubprt{\NPPS}{170}{248}{2007}
[arXiv:hep-ex/0702012].

\bibitem{NPPS187p145}
B.~D.~Yabsley,
{\it Hidden and open charm at Belle (and elsewhere)},
\pubprt{\NPPS}{187}{145}{2009}
[arXiv:0811.0873 [hep-ex]].

\bibitem{PLB625p212}
S.~L.~Zhu,
{\it The possible interpretations of Y(4260)},
\pubprt{\PLB}{625}{212}{2005}
[arXiv:hep-ph/0507025].

\bibitem{PRD72p031503}
F.~J.~Llanes-Estrada,
{\it Y(4260) and possible charmonium assignment},
\pubprt{\PRD}{72}{031503}{2005}
[arXiv:hep-ph/0507035].

\bibitem{PRD72p054023}
X.~Liu, X.~Q.~Zeng and X.~Q.~Li,
{\it Possible molecular structure of the newly observed Y(4260)},
\pubprt{\PRD}{72}{054023}{2005}
[arXiv:hep-ph/0507177].

\bibitem{PLB631p164}
E.~Kou and O.~Pene,
{\it Suppressed decay into open charm for the Y(4260) being an hybrid},
\pubprt{\PLB}{631}{164}{2005}
[arXiv:hep-ph/0507119].

\bibitem{PRD72p031502}
L.~Maiani, V.~Riquer, F.~Piccinini and A.~D.~Polosa,
{\it Four quark interpretation of Y(4260)},
\pubprt{\PRD}{72}{031502}{2005}
[arXiv:hep-ph/0507062].

\bibitem{HEPEX0701002}
D.~V.~Bugg,
{\it  The $\pi\pi$ mass spectrum in $Y(4260)\to\pi\pi J/\psi$},
arXiv:hep-ex/0701002.

\bibitem{PRD76p111105}
B.~Aubert  [BaBar Collaboration],
{\it Study of the exclusive initial-state radiation production
of the $D\bar{D}$ system},
\pubprt{\PRD}{76}{111105}{2007}.

\bibitem{ARXIV07100165}
B.~W.~Lang,
{\it Charm-production in $e^{+}e^{-}$  annihilation around 4 GeV},
Published in the Proceedings of International Workshop on Charm Physics
(Charm 2007), Ithaca, New York, 5-8 Aug 2007, 1,
[arXiv:0710.0165 [hep-ex]].

\bibitem{PLB69p503}
G.~Goldhaber {\it et al.},
{\it $D$ and $D^{\ast}$ meson production near 4-Gev in
$e^{+}e^{-}$ annihilation},
\pubprt{\PLB}{69}{503}{1977}.

\bibitem{NPB471p59}
D.~V.~Bugg, B.~S.~Zou and A.~V.~Sarantsev,
{\it New results on $\pi\pi$ phase shifts between 600-MeV and 1900-MeV},
Nucl.\ Phys.\  B {\bf 471}, 59 (1996).

\bibitem{ZPC68p647}
N.~A.~T\"{o}rnqvist,
{\it Understanding the scalar meson nonet},
\pubprt{\ZPC}{68}{647}{1995}
[arXiv:hep-ph/9504372].

\bibitem{ZPC13p43}
S.~B.~Gerasimov and A.~B.~Govorkov,
{\it Radial excitations of $\rho^{-}$ and $\pi$ mesons
and their strong decays},
\pubprt{\ZPC}{13}{43}{1982}.

\bibitem{PRD79p014001}
G.~J.~Ding,
{\it Are $Y(4260)$ and $Z_{2}^{+}$(4250) $D_{1}D$ or $D_{0}D^{\ast}$
hadronic molecules?},
\pubprt{\PRD}{79}{014001}{2009}
[arXiv:0809.4818 [hep-ph]].

\bibitem{ARXIV08111755}
E.~van Beveren and G.~Rupp,
{\it The spectrum of charmonium in the Resonance-Spectrum Expansion},
in Proceedings {\it Bled Workshops in Physics},
Vol.~9, no.~1, pp 26-29 (2008)
[arXiv:0811.1755 [hep-ph]].

\bibitem{PRD56p4062}
A.~E.~Dorokhov, S.~V.~Esaibegian and S.~V.~Mikhailov,
{\it Virtualities of quarks and gluons in QCD vacuum and nonlocal
condensates within single instanton approximation},
\pubprt{\PRD}{56}{4062}{1997}
[arXiv:hep-ph/9702417].

\bibitem{BJP34p865}
L.~A.~Trevisan, A.~E.~Dorokhov and L.~Tomio,
{\it Path dependence of the quark nonlocal condensate
within the instanton model},
\pubprt{\BJP}{34}{865}{2004}
[arXiv:hep-ph/0405293].

\bibitem{JPG14p1037}
A.~Buchmann, Y.~Yamauchi, H.~Ito and A.~Faessler,
{\it  The pion pair current in the quark cluster model},
\pubprt{\JPG}{14}{1037}{1988}.

\bibitem{PLB495p300}
E.~van Beveren, G.~Rupp and M.~D.~Scadron,
{\it Why is the $f_{0}(9890)$ is mostly $s\bar{s}$},
\pubprt{\PLB}{495}{300}{2000}
\ertbid{\ B}{509}{365}{2001}
[arXiv:hep-ph/0009265].

\bibitem{NPB320p1}
J.~E.~Augustin {\it et al.}  [DM2 Collaboration],
{\it Study of the $J/\Psi$ decay into five pions},
\pubprt{\NPB}{320}{1}{1989}.

\bibitem{PRL53p2465}
H.~Aihara {\it et al.},
{\it Evidence for the $F^{\ast}$ meson},
\pubprt{\PRL}{53}{2465}{1984}.

\bibitem{PLB156p441}
A.~E.~Asratyan {\it et al.},
{\it Charmed strange vector meson production in
antineutrino-nucleon interactions},
\pubprt{\PLB}{156}{441}{1985}.

\bibitem{PRL58p2171}
G.~T.~Blaylock {\it et al.}  [Mark III Collaboration],
{\it Observation of $e^{+}e^{-}\to D_{s}^{\pm}D_{s}^{\ast\mp}$
at $\sqrt{s}=4.14$ GeV},
\pubprt{\PRL}{58}{2171}{1987}.

\bibitem{PLB207p349}
H.~Albrecht {\it et al.}  [ARGUS Collaboration],
{\it Measurement of $D_{s}^{\ast}-D_{s}$ mass difference},
\pubprt{\PLB}{207}{349}{1988}.

\bibitem{PRD50p1884}
D.~Brown {\it et al.}  [CLEO Collaboration],
{\it Precision measurement of $D_{s}^{\ast +}-D_{s}^{+}$ mass difference},
\pubprt{\PRD}{50}{1884}{1994}.

\bibitem{PRL75p3232}
J.~Gronberg {\it et al.}  [CLEO Collaboration],
{\it Observation of the isospin violating decay
$D_{s}^{\ast +}\to D_{s}^{+}\pi^{0}$},
\pubprt{\PRL}{75}{3232}{1995}.

\bibitem{PRD27p1527}
E.~van Beveren, G.~Rupp, T.~A.~Rij\-ken, and C.~Dullemond,
{\it Radial spectra and hadronic decay widths of light and heavy mesons},
\pubprt{\PRD}{27}{1527}{1983}.

\bibitem{HEPPH0201006}
E.~van Beveren and G.~Rupp,
{\it Scalar mesons within a model for all non-exotic mesons},
in Proc.\ Workshop {\it Recent Developments in Particle
and Nuclear Physics, April 30, 2001, Coimbra (Portugal)},
(Universidade de Coimbra, 2003) ISBN 972-95630-3-9, pages 1--16,
arXiv:hep-ph/0201006.

\bibitem{ARXIV08091149}
E.~van~Beveren, and G.~Rupp,
{\it Meson-meson interactions and Regge propagators},
arXiv:0809.1149 [hep-ph].

\bibitem{JPG35p075005}
D.~V.~Bugg,
{\it How resonances can synchronise with thresholds},
\pubprt{\JPG}{35}{075005}{2008}
[arXiv:0802.0934 [hep-ph]].

\bibitem{EKA}
K.~P.~Khemchandani, A.~Martinez~Torres and E.~Oset,
in preparation.

\bibitem{ARXIV09032229}
T.~Aaltonen {\it et al.}  [The CDF collaboration],
{\it Evidence for a narrow near-threshold structure in the $J/\psi\phi$
mass spectrum in $B^{+}\to J/\psi\phi K^{+}$ decays},
arXiv:0903.2229 [hep-ex].
\end{thebibliography}
\end{document}